\documentclass[hyperref,UTF8]{ctexart}
\usepackage{CJKutf8}
\usepackage{CJK}
\usepackage{geometry}
\usepackage{cite}
\usepackage{latexsym,bm,amsmath,amssymb,mathrsfs}
\usepackage{stmaryrd} 
\usepackage{mathrsfs}
\usepackage{mathtools} 
\usepackage{threeparttable} 
\newtheorem{definition}{Definition}[section]
\newtheorem{theorem}[definition]{Theorem}

\newtheorem{lemma}[definition]{Lemma} 

\usepackage{caption}
\usepackage[figuresright]{rotating} 

\usepackage{enumerate}
\usepackage{url}
\usepackage{datetime} 
\usepackage{algorithm}
\usepackage{algorithmic}
\usepackage{verbatim} 
\usepackage{setspace}
\usepackage{float} 
\usepackage{subfigure} 
\usepackage{mathrsfs}
\usepackage{bm}
\usepackage{makecell}
\CTEXsetup[format={\Large\bfseries}]{section} 
\floatname{algorithm}{Algorithm}
\usepackage{multirow}
\usepackage{booktabs}
\usepackage{color} 

\usepackage[numbers,sort&compress]{natbib}

\usepackage{threeparttable} 

\begin{document}
\begin{CJK*}{UTF8}{}
\title{\textbf{\Large{Multi-mode Tensor Train Factorization with Spatial-spectral Regularization for Remote Sensing Images Recovery}}}
\author{Gaohang Yu\footnotemark[1],\  \ Shaochun Wan\footnotemark[1],\ \ Liqun Qi\footnotemark[2],\ \ Yanwei Xu\footnotemark[3]}

\footnotetext[1]{Department of Mathematics, Hangzhou Dianzi University, 310018, China. E-mail: maghyu@hdu.edu.cn}
\footnotetext[2]{Department of Applied Mathematics, The Hong Kong Polytechnic University, Hung Hom, Kowloon, Hong Kong. E-mail: liqun.qi@polyu.edu.hk}
\footnotetext[3]{Huawei Theory Research Lab, Hong Kong, China. E-mail: xuyanwei1@huawei.com}
\date{}
\maketitle
\numberwithin{equation}{section}
\setcounter{page}{1}
\noindent\textbf{Abstract: }Tensor train (TT) factorization and corresponding TT rank, which can well express the low-rankness and mode correlations of higher-order tensors, have attracted much attention in recent years. However, TT factorization based methods are generally not sufficient to characterize low-rankness along each mode of third-order tensor. Inspired by this, we generalize the tensor train factorization to the mode-k tensor train factorization and introduce a corresponding multi-mode tensor train (MTT) rank. Then, we proposed a novel low-MTT-rank tensor completion model via  multi-mode TT factorization and spatial-spectral smoothness regularization. To tackle the proposed model, we develop an efficient proximal alternating minimization (PAM) algorithm. Extensive numerical experiment results on visual data demonstrate that the proposed MTTD3R method outperforms compared methods in terms of visual and quantitative measures.
\vspace{5mm}
\section{Introduction}
\setlength{\baselineskip}{20pt}
With the rapid development of information technology, realistic data, such as remote sensing image, color image and video, tend to have high dimensions and complex structures. Tensor, as a high-dimensional generalization of vector and matrix, can better characterize the complex essential structures of higher-order data. Besides, higher-order tensors have extensive applications in many fields, such as multispectral image (MSI) recovery \cite{zhao2020deep}, hyperspectral image (HSI) restoration \cite{chen2018hyperspectral,zheng2019mixed,Xu2013Parallel}, image/video inpainting \cite{Xu2013Parallel,Liu2013Tensor,Zhou2018Tensor,Zhang2014Novel,li2017low}, and signal reconstruction \cite{ding2019total}.\par

Due to unacceptable cost of collecting complete data or loss of information during transmission, many real-world higher-order tensor data may contain missing entries. Therefore, \emph{tensor completion}, which estimates the values of missing tensor entries, becomes one of the most important problems in tensor analysis and processing. Fortunately, many multi-dimensional tensors, such as color images and hyperspectral images (HSIs), are intrinsically or approximately low-rank and contain wealthy spatial-spectral information. As an extension of low-rank matrix completion (LRMC) \cite{Candes2009Exact}, low-rank tensor completion (LRTC) utilizes the low-rank prior to express the relationship between the observed and missing entries, and can be mathematically written as:
\begin{equation}\label{equation:1.1}
		\begin{aligned}
			\mathop {\min }\limits_{ \mathcal{C}} \quad & \mbox{rank}(\mathcal{C}) , \\
			\mbox{s.t.} \quad & {\mathcal{P}_\Omega }({\cal C} - {\cal M}) = 0,
		\end{aligned}
	\end{equation}
where $\mathcal{C}$ is the underlying tensor, $\mathcal{M}$ is the observed tensor, $\Omega$ is the index set for available entries, and $\mathcal{P}_{\Omega}(\cdot)$ is the projection operator that keeps the entries of $\mathcal{C}$ in $\Omega$ and zeros out others.\par
J. Liu et al. \cite{Liu2013Tensor} generalized matrix trace norm to tensor case and proposed a tensor completion model based on the defined tensor trace norm:
\begin{equation}\label{equation:1}
    \begin{aligned}
        \min_{\mathcal{X}}\quad &\sum_{n=1}^N \alpha_n\Vert C_{(n)}\Vert_{\ast},\\
        \mbox{s.t.}\quad & \mathcal{P}_{\Omega}(\mathcal{C})=\mathcal{P}_{\Omega}(\mathcal{M}).
    \end{aligned}
\end{equation}
where $\alpha_n\geq 0$ and satisfies $\sum\nolimits_{n=1}^N \alpha_n=1$, $C_{(n)}$ is the mode-$n$ unfolding matrix of $\mathcal{C}$. Besides, three algorithms (SiLRTC, FaLRTC, HaLRTC) are developed to solve model (\ref{equation:1}). Then, N. Liu \cite{liu2021hyperspectral} et al. utilized the tensor trace norm as a convex surrogate for rank and proposed a low-rank tensor approximation (LRTA) model for hyperspectral and multispectral (HS-MS) fusion.\par
However, all these trace norm minimization methods involve the singular value decomposition (SVD) of $C_{(n)}$ and thus suffer from high computational cost. To tackle this issue, Xu et al. \cite{Xu2013Parallel} utilized the matrix factorization method to preserve the low-rank structure of each mode matricization for efficiently handling large scale unfolding matrices:
\begin{equation}\label{equation:2}
		\begin{aligned}
			\mathop {\min }\limits_{X, Y, \mathcal{C}} \quad & \sum\limits_{n=1}^N {\dfrac{\alpha _n}{2}{{\left\| {{X_n}{Y_n} - {C_{(n)}}} \right\|}_F^2}} , \\
			\mbox{s.t.} \quad & {\mathcal{P}_\Omega }({\cal C} - {\cal M}) = 0,
		\end{aligned}
	\end{equation}
where $X=(X_1, \ldots, X_N)$, $Y=(Y_1, \ldots, Y_N)$, $\alpha_n$ are constants satisfying $\alpha_n\geq 0$ and $\sum\limits_{n=1}^N \alpha_n =1$. The method, which is named as low-rank tensor completion by parallel matrix factorization (TMac), has shown better performance than FaLRTC. Many real-world data exhibits piecewise smooth prior, however, Xu et al. only consider the low-rank prior. Due to the ability of total variation (TV) to preserve edges\cite{rudin1992nonlinear}, Ji et al. \cite{ji2016tensor} considered to introduce it into the tensor completion problem (\ref{equation:2}).
\begin{equation}\label{equation:2.11}
		\begin{aligned}
			\mathop {\min }\limits_{\mathcal{C}, A, X} \quad & \sum\limits_{n=1}^N {\dfrac{\alpha _n}{2}{{\left\| {C_{(n)}-A_n X_n} \right\|}_F^2}} +\mu\mbox{TV}(X_3), \\
			\mbox{s.t.} \quad & {\mathcal{P}_\Omega }({\cal C} - {\cal M}) = 0,
		\end{aligned}
	\end{equation}
where $\mu$ is the regularization parameter, $A=(A_1,\ldots,A_N)$, $X=(X_1,\ldots,X_N)$, and $\mbox{TV}(X_3)$ is the total variation of $X_3$.\par
As pointed out by \cite{kilmer2013third,kilmer2011factorization}, directly unfolding a tensor would destroy the original multi-way structure of the data, leading to vital information loss and degraded performance. Recently, on the basis of tensor-tensor product (t-product) and tensor singular value decomposition (t-SVD) \cite{kilmer2011factorization,Oseledets2011}, Kilmer et al. \cite{kilmer2013third} introduced the definitions of tensor multi-rank and tubal rank. Afterward, Semerci et al. \cite{Semerci2014Tensor} developed a new tensor nuclear norm (TNN) to better preserve the inherent low-rank structure of tensor. Then, Zhang et al. \cite{Zhang2014Novel} proposed a TNN-based low-rank tensor completion model and applied it to video inpainting. It's worth noting that TNN-based methods still involve computing t-SVD and thus time-consuming. Zhou et al. \cite{Zhou2018Tensor} proposed a low-tubal-rank tensor factorization model (TCTF) to avoid computing t-SVD, which factorizes the target tensor into the t-product of two smaller tensors:
\begin{equation}\label{equation:3}
		\begin{aligned}
			\mathop {\min }\limits_{{\cal X},{\cal Y},{\cal C}} \quad & \frac{1}{2}\left\| {{\cal X} * {\cal Y} - {\cal C}} \right\|_F^2\\
			\mbox{s.t.} \quad  & {{\cal P}_\Omega }({\cal C} - {\cal M}) = 0.
		\end{aligned}
	\end{equation}
\indent In our recent work, we showed that the TCTF method lacks characterization of mode correlations. As pointed in \cite{bengua2017efficient,zhang2021multiscale}, tensor train (TT) factorization is especially suitable for high-
dimensional tensors and can better characterize the global mode correlations. Thus, we proposed a streaming tensor completion method TTD2R on the basis of tensor train factorization and spatial-temporal constraint:
\begin{equation}\label{eq:3.11}
\begin{aligned}
\mathop {\min }\limits_{{\cal C},{\cal X},{\cal Y},{\cal Z}}\quad & \dfrac{1}{2}\left\|{\cal C}-\llbracket{\cal X};{\cal Y};{\cal Z}\rrbracket\right\|_{F}^2+\dfrac{\alpha_1}{2}\left\|\mathcal{D}_h(\mathcal{C})\right\|_{F}^2+\dfrac{\alpha_2}{2}\left\|\mathcal{D}_v(\mathcal{C})\right\|_{F}^2,\\
\mbox{s.t.}\quad\quad & P_\Omega(\mathcal{C})=P_\Omega(\mathcal{M}),
\end{aligned}
\end{equation}
where $\llbracket{\cal X};{\cal Y};{\cal Z}\rrbracket$ denotes TT factorization of $\mathcal{C}\in\mathbb{R}^{I_1\times I_2\times I_3}$, $\mathcal{X},\mathcal{Y},\mathcal{Z}$ are TT cores, $\alpha_1$ and $\alpha_2$ are smoothness regularization parameters, $\mathcal{D}_h(\cdot)$ and $\mathcal{D}_v(\cdot)$ are first-order difference operators. TTD2R method outperforms other compared methods on color images inpainting and traffic data recovery, but it's still not sufficient to express tensor data with low-rank property and smoothness along each mode.\par
In this paper, we define the mode-k tensor train factorization and MTT rank, which can more flexibly and accurately characterize low-rankness of HSIs, MSIs and gray videos. After that, we propose a novel low-rank tensor completion model for visual data recovery, which integrates MTT factorization with spatial-spectral smoothness.\par
The outline of this paper is as follows. Section \ref{section:2} summarizes some notations, designs the mode-k tensor train factorization, and defines the multi-mode tensor train rank. Section \ref{section:3} proposes a MTT factorization based third-order tensor completion model with spatial-spectral smoothness regularization terms, then a theoretically and numerically convergent PAM-based algorithm is developed to solve the problem. Section \ref{section:5} evaluates the performance of the proposed method on color images, gray videos, MSIs, and remote sensing HSIs. Section \ref{section:6} concludes this article.

\section{Preliminaries}\label{section:2}
\subsection{Notations}
We denote vectors as bold lowercase letters (e.g., $\boldsymbol{a}$), matrices as uppercase letters (e.g., A), and tensors as calligraphic letters (e.g., $\mathcal{A}$). For a three-way tensor $\mathcal{A}\in\mathbb{R}^{I_1\times I_2\times I_3}$, following the MATLAB notation, we denote its ($i, j, s$)th element as $\mathcal{A}(i, j,s)$, its ($i,j$)th mode-1, mode-2, and mode-3 fibers as $\mathcal{A}(:,i, j)$, $\mathcal{A}(i,:, j)$, and $\mathcal{A}(i, j,:)$, respectively. For the sake of clarity, we use $A_1^{(i)}\in\mathbb{R}^{I_3\times I_2}$, $A_2^{(i)}\in\mathbb{R}^{I_1\times I_3}$, and $A_3^{(i)}\in\mathbb{R}^{I_2\times I_1}$ to denote the $i$th mode-1 (horizontal), mode-2 (lateral), and mode-3 (frontal) slices of $\mathcal{A}$, respectively. The Frobenius norm of $\mathcal{A}$ is defined as  $\Vert\mathcal{A}\Vert_F\coloneqq\sqrt{\sum_{i,j,s}\vert\mathcal{A}(i,j,s)\vert^2}$. The $\ell_1$ norm of $\mathcal{A}$ is defined as $\Vert\mathcal{A}\Vert_1\coloneqq \sum_{i,j,s}\vert\mathcal{A}(i,j,s)\vert$. Denote the mode-$k$ matricization of $\mathcal{A}$ as $A_{(k)}$.

\subsection{Multi-mode Tensor Train Factorization and Corresponding Ranks}
\begin{definition}{(Tensor Mode-$k$ Permutation, refer to the definition in \cite{zheng2019mixed}):}
  Given a three-way tensor $\mathcal{A}\in\mathbb{R}^{I_1\times I_2\times I_3}$. The mode-$k$ permutation of $\mathcal{A}$, denoted by $\vec{\mathcal{A}}^k$, is defined as the tensor whose $i$th mode-$2$ slice is the $i$th mode-$k$ slice of $\mathcal{A}$., i.e., $\mathcal{A}(i,j,s)=\vec{\mathcal{A}}^1(s,i,j)=\vec{\mathcal{A}}^2(i,j,s)=\vec{\mathcal{A}}^3(j,s,i)$. We define the corresponding operation as $\vec{\mathcal{A}}^k\coloneqq\text{permute}(\mathcal{A},k)$ and its inverse operation as $\mathcal{A}\coloneqq\text{ipermute}(\vec{\mathcal{A}}^k,k)$.
\end{definition}
\begin{figure}[H]
  \centering
  \includegraphics[width=0.8\linewidth]{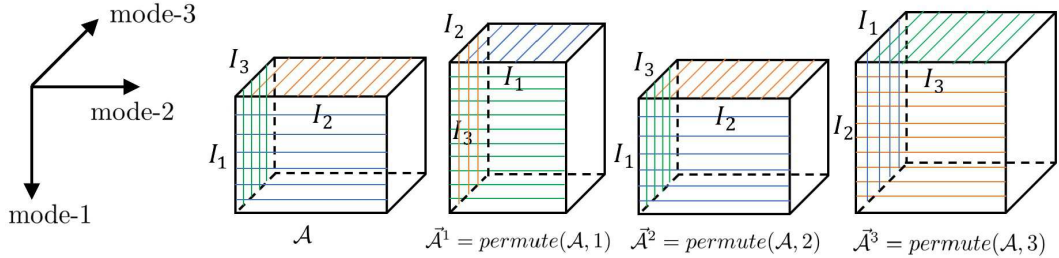}\\
  \caption{Illustration of the mode-$k$ permutation for an $I_1\times I_2\times I_3$ tensor.}\label{fig:1}
\end{figure}
\par Figure \ref{fig:1} shows the mode-$k$ permutation of an $I_1\times I_2\times I_3$ tensor. With the above definitions, we design the mode-$k$ tensor train factorization of a third-order tensor.

\begin{definition}{(Mode-$k$ Tensor Train Factorization):}
  The mode-$1$ tensor train (TT) factorization of $\mathcal{A}\in\mathbb{R}^{I_1\times I_2\times I_3}$ is defined as
    \begin{equation*}
    \mathcal{A}(i_1,i_2,i_3)=\mathcal{X}_1^{(i_3)} \mathcal{Y}_1^{(i_1)} \mathcal{Z}_1^{(i_2)}
  \end{equation*}
  where $\mathcal{X}_1\in\mathbb{R}^{I_3\times r_1^1\times 1}$, $\mathcal{Y}_1\in\mathbb{R}^{I_1\times r_2^1\times r_1^1}$, $\mathcal{Z}_1\in\mathbb{R}^{I_2\times 1\times r_2^1}$, $(r_1^1,r_2^1)\coloneqq (\mbox{rank}(A_{(3)}),\mbox{rank}(A_{(2)}))$ is the mode-$1$ TT rank of $\mathcal{A}$. Denote the mode-1 TT factorization by $\mathcal{A}=\llbracket\mathcal{X}_1;\mathcal{Y}_1;\mathcal{Z}_1\rrbracket_1$ for brevity.\par
  The mode-$2$ TT factorization of $\mathcal{A}\in\mathbb{R}^{I_1\times I_2\times I_3}$ is defined as
    \begin{equation*}
    \mathcal{A}(i_1,i_2,i_3)=\mathcal{X}_2^{(i_1)} \mathcal{Y}_2^{(i_2)} \mathcal{Z}_2^{(i_3)}
  \end{equation*}
  where $\mathcal{X}_2\in\mathbb{R}^{1\times I_1\times r_1^2}$, $\mathcal{Y}_2\in\mathbb{R}^{r_1^2\times I_2\times r_2^2}$, $\mathcal{Z}_2\in\mathbb{R}^{r_2^2\times I_3\times 1}$, $(r_1^2,r_2^2)\coloneqq (\mbox{rank}(A_{(1)}),\mbox{rank}(A_{(3)}))$ is the mode-$2$ TT rank of $\mathcal{A}$. Denote it by $\mathcal{A}=\llbracket\mathcal{X}_2;\mathcal{Y}_2;\mathcal{Z}_2\rrbracket_2$.\par
  The mode-$3$ TT factorization of $\mathcal{A}\in\mathbb{R}^{I_1\times I_2\times I_3}$ is defined as
    \begin{equation*}
    \mathcal{A}(i_1,i_2,i_3)=\mathcal{X}_3^{(i_2)} \mathcal{Y}_3^{(i_3)} \mathcal{Z}_3^{(i_1)}
  \end{equation*}
  where $\mathcal{X}_3\in\mathbb{R}^{r_1^3\times 1\times I_2}$, $\mathcal{Y}_3\in\mathbb{R}^{r_2^3\times r_1^3\times I_3}$, $\mathcal{Z}_3\in\mathbb{R}^{1\times r_2^3\times I_1}$, $(r_1^3,r_2^3)\coloneqq (\mbox{rank}(A_{(2)}),\mbox{rank}(A_{(1)}))$ is the mode-$3$ TT rank of $\mathcal{A}$. Denote it by $\mathcal{A}=\llbracket\mathcal{X}_3;\mathcal{Y}_3;\mathcal{Z}_3\rrbracket_3$.
\end{definition}
Actually, from above definitions, it's easy to prove that $\mathcal{A}=\llbracket\mathcal{X}_k;\mathcal{Y}_k;\mathcal{Z}_k\rrbracket_k$ if and only if $\vec{\mathcal{A}}^k=\llbracket\vec{\mathcal{X}_k}^k;\vec{\mathcal{Y}_k}^k;\vec{\mathcal{Z}_k}^k\rrbracket_2$. That is, the mode-$k$ TT ranks of $\mathcal{A}$ is equal to the mode-2 TT ranks of $\vec{\mathcal{A}}^k$.
\begin{definition}{(Multi-mode Tensor Train Rank)}
  The multi-mode tensor train (MTT) rank of a third-order tensor $\mathcal{A}\in\mathbb{R}^{I_1\times I_2\times I_3}$, denoted as $\rm{rank}_{\rm{MTT}}(\mathcal{A})$, is defined as a vector, whose $k$th element is the mode-$k$ TT rank.
\end{definition}

\section{Tensor completion combining multi-mode tensor train factorization and spatial-spectral regularization}\label{section:3}
We establish the following low-rank tensor completion model by minimizing the MTT rank $\mbox{rank}_{\mbox{MTT}}(\cdot)$ of underlying tensor:

\begin{equation}\label{eq:3.1}
    \begin{aligned}
        \min_{\mathcal{A}\in\mathbb{R}^{I_1\times I_2\times I_3}}\quad & \rm{rank}_{MTT}(\mathcal{A})\\
        \mbox{s.t.}\quad\quad & P_\Omega(\mathcal{A})=P_\Omega(\mathcal{M})
    \end{aligned}
\end{equation}
\noindent where $\mathcal{A}\in\mathbb{R}^{I_1\times I_2\times I_3}$ is the tensor to be filled; $\mathcal{M}$ is a normalized observed tensor such that $0\leq\mathcal{M}\leq 1$.
\par In this paper, we utilize multi-mode tensor train factorization to flexibly characterize low-rankness along each mode of third-order tensor:
\begin{equation}\label{eq:3.2}
    \begin{aligned}
        \mathop{\min }\limits_{{\cal A},{\cal X}_u,{\cal Y}_u,{\cal Z}_u}\quad & \sum_{u=1}^3 \frac{\alpha_u}{2}\left\| {{\cal A} - \llbracket{\cal X}_u; {\cal Y}_u; {\cal Z}_u}\rrbracket_u \right\|_F^2,\\
        \mbox{s.t.}\quad\quad & P_\Omega(\mathcal{A})=P_\Omega(\mathcal{M}),
    \end{aligned}
\end{equation}
where $ \mathcal{X}_u $,$ \mathcal{Y}_u $,$ \mathcal{Z}_u $ are cores of the mode-$u$ TT factorization of $\mathcal{A}$.

\par The low-MTT-rank constraint in (\ref{eq:3.2}) can be interpreted as a global feature, as it concerns the construction of the whole tensor. However, the low-rankness itself is generally not sufficient to recover the underlying data. As another significant prior, spatial-spectral smoothness appears in many real-world multi-dimensional data. To get better completion performance, we incorporate spatial-spectral regularization into (\ref{eq:3.2}) and obtain the improved version:


\begin{equation}\label{eq:3.3}
\begin{aligned}
\mathop{\min }\limits_{{\cal A},{\cal X}_u,{\cal Y}_u,{\cal Z}_u}\quad & \sum_{u=1}^3 \frac{\alpha_u}{2}\left\| {{\cal A} - \llbracket{\cal X}_u; {\cal Y}_u; {\cal Z}_u}\rrbracket_u \right\|_F^2+\frac{\mu}{2}\Vert\mathcal{D}_w(\mathcal{A})\Vert_F^2,\\
\mbox{s.t.}\quad\quad & P_\Omega(\mathcal{A})=P_\Omega(\mathcal{M}),
\end{aligned}
\end{equation}

\noindent where $\mathcal{D}_w(\cdot)=\left[w_1\times\mathcal{D}_h(\cdot);w_2\times\mathcal{D}_v(\cdot);w_3\times\mathcal{D}_t(\cdot)\right]$ is the so-called weighted three-dimensional difference operator \cite{wang2017hyperspectral,wang2022total} and $\mathcal{D}_h(\cdot)$, $\mathcal{D}_v(\cdot)$, $\mathcal{D}_t(\cdot)$ are the first-order difference operators with respect to three different directions of the tensor cube.
\begin{equation*}
  \mathcal{D}_h(\mathcal{A}(i,j,s))=
  \begin{cases}
    \mathcal{A}(i+1,j,s)-\mathcal{A}(i,j,s) &\mbox{if $i\neq I_1$}\\
    \mathcal{A}(1,j,s)-\mathcal{A}(i,j,s) &\mbox{if $i=I_1$}
  \end{cases}
\end{equation*}
\begin{equation*}
  \mathcal{D}_v(\mathcal{A}(i,j,s))=
  \begin{cases}
    \mathcal{A}(i,j+1,s)-\mathcal{A}(i,j,s) &\mbox{if $i\neq I_2$}\\
    \mathcal{A}(i,1,s)-\mathcal{A}(i,j,s) &\mbox{if $i=I_2$}
  \end{cases}
\end{equation*}
\begin{equation*}
  \mathcal{D}_t(\mathcal{A}(i,j,s))=
  \begin{cases}
    \mathcal{A}(i,j,s+1)-\mathcal{A}(i,j,s) &\mbox{if $i\neq I_3$}\\
    \mathcal{A}(i,j,1)-\mathcal{A}(i,j,s) &\mbox{if $i=I_3$}
  \end{cases}
\end{equation*}
Moreover, problem (\ref{eq:3.3}) can be rewritten as
\begin{equation}\label{eq:3.4}
\mathop {\min }\limits_{{\cal A},{\cal X}_u,{\cal Y}_u,{\cal Z}_u}\quad  f(\mathcal{X}_u,\mathcal{Y}_u,\mathcal{Z}_u,\mathcal{A})
\end{equation}
where
\begin{equation*}
  f(\mathcal{X}_u,\mathcal{Y}_u,\mathcal{Z}_u,\mathcal{A}) = f_1(\mathcal{X}_1,\mathcal{Y}_1,\mathcal{Z}_1,\mathcal{A})+f_2(\mathcal{X}_2,\mathcal{Y}_2,\mathcal{Z}_2,\mathcal{A})+
  f_3(\mathcal{X}_3,\mathcal{Y}_3,\mathcal{Z}_3,\mathcal{A})+g(\mathcal{A})+\delta_S(\mathcal{A}),
\end{equation*}
\begin{equation*}
  f_1(\mathcal{X}_1,\mathcal{Y}_1,\mathcal{Z}_1,\mathcal{A})=\frac{\alpha_1}{2}\left\| {{\cal A} - \llbracket{\cal X}_1; {\cal Y}_1; {\cal Z}_1}\rrbracket_1 \right\|_F^2,
\end{equation*}
\begin{equation*}
  f_2(\mathcal{X}_2,\mathcal{Y}_2,\mathcal{Z}_2,\mathcal{A})=\frac{\alpha_2}{2}\left\| {{\cal A} - \llbracket{\cal X}_2; {\cal Y}_2; {\cal Z}_2}\rrbracket_2 \right\|_F^2,
\end{equation*}
\begin{equation*}
  f_3(\mathcal{X}_3,\mathcal{Y}_3,\mathcal{Z}_3,\mathcal{A})=\frac{\alpha_3}{2}\left\| {{\cal A} - \llbracket{\cal X}_3; {\cal Y}_3; {\cal Z}_3}\rrbracket_3 \right\|_F^2,
\end{equation*}
\begin{equation*}
  g(\mathcal{A})=\frac{\mu}{2}\Vert\mathcal{D}_w(\mathcal{A})\Vert_F^2,
\end{equation*}
\begin{equation*}
  S=\left\{\mathcal{W}\in\mathbb{R}^{I_1\times I_2\times I_3}\big|\mathcal{W}(i, j, k)=\mathcal{M}(i,j,k)\ \mbox{for}\ (i,j,k)\in\Omega; \ \mbox{while}\ \mathcal{W}(i, j, k)\in\left[0, 1\right] \ \mbox{for}\ (i, j, k) \notin\Omega\right\},
\end{equation*}
\begin{equation*}
  \delta_S\left(\mathcal{A}\right)=\left\{
    \begin{array}{ll}
      0, & \mathcal{A}\in S; \\
      +\infty, & \mathcal{A}\notin S.
    \end{array}
  \right.
\end{equation*}

\section{Algorithm for solving (\ref{eq:3.4})}\label{section:4}
\subsection{Algorithm description}
Model (\ref{eq:3.4}) is a multivariate optimization problem. Alternate Minimization(AM)  algorithm is usually used to solve multivariate optimization problems due to its simplicity and efficiency. To improve the theoretical convergence and numerical stability of the AM algorithm, proximal terms are suggested to add in subproblems generated by AM algorithm\cite{attouch2010proximal,lin2020tensor}, which is called the Proximal Alternate Minimization (PAM) algorithm.\par
Given the initial point $(\mathcal{X}_{u,k},\mathcal{Y}_{u,k},\mathcal{Z}_{u,k},\mathcal{A}_k)$ of the problem (\ref{eq:3.4}), the PAM iteration is defined as follows:
\begin{equation}\label{eq:4.1}
\mathcal{X}_{u,k+1}=\mathop {\arg \min }\limits_{\mathcal{X}_u} \ f_u({\cal X}_u,{\cal Y}_{u,k},{\cal Z}_{u,k},{\cal A}_k)+\frac{\rho}{2}\Vert\mathcal{X}_u-\mathcal{X}_{u,k}\Vert_F^2,\ u\in\Xi(3),
\end{equation}
\begin{equation}\label{eq:4.2}
\mathcal{Y}_{u,k+1}=\mathop {\arg \min }\limits_{\mathcal{Y}_u} \ f_u({\cal X}_{u,k+1},{\cal Y}_{u},{\cal Z}_{u,k},{\cal A}_k)+\frac{\rho}{2}\Vert\mathcal{Y}_u-\mathcal{Y}_{u,k}\Vert_F^2,\ u\in\Xi(3),
\end{equation}
\begin{equation}\label{eq:4.3}
\mathcal{Z}_{u,k+1}=\mathop {\arg \min }\limits_{\mathcal{Z}_u} \ f_u({\cal X}_{u,k+1},{\cal Y}_{u,k+1},{\cal Z}_{u},{\cal A}_k)+\frac{\rho}{2}\Vert\mathcal{Z}_u-\mathcal{Z}_{u,k}\Vert_F^2,\ u\in\Xi(3),
\end{equation}
\begin{equation}\label{eq:4.4}
\mathcal{A}_{k+1}=\mathop {\arg \min }\limits_{\mathcal{A}} \ f(\boldsymbol{x}_{k+1},\boldsymbol{y}_{k+1},\boldsymbol{z}_{k+1},{\cal A})+\frac{\rho}{2}\Vert\mathcal{A}-\mathcal{A}_{k}\Vert_F^2,
\end{equation}

\noindent where $\boldsymbol{x}_{k+1} = ({\cal X}_{1,k+1}; {\cal X}_{2,k+1}; {\cal X}_{3,k+1})$, $\boldsymbol{y}_{k+1} = ({\cal Y}_{1,k+1}; {\cal Y}_{2,k+1}; {\cal Y}_{3,k+1})$, $\boldsymbol{z}_{k+1} = ({\cal Z}_{1,k+1}; {\cal Z}_{2,k+1}; {\cal Z}_{3,k+1})$, $\rho>0$ is the given parameter.
\par It is obvious that  (\ref{eq:4.1})-(\ref{eq:4.4}) are all strongly convex optimization problems, whose existence and uniqueness are guaranteed. In Subsection \ref{Sec:4.2}, efficient methods will be introduced to solve ({\ref{eq:4.1}})-(\ref{eq:4.4}).

\subsection{Algorithm Implementation}\label{Sec:4.2}
In this section, we will concentrate on solving subproblems (\ref{eq:4.1})-(\ref{eq:4.4}) arising from the PAM algorithm.
\par Firstly, it is easy to check that (\ref{eq:4.1}) and (\ref{eq:4.3}) have an unique closed-form solution respectively as follows:
\begin{equation}\label{eq:4.5}
{X}_{u,k+1} = \bigg[\rho I+\alpha_u\sum_{i=1}^{I_u}(Y_{u,k}^{(i)} Z_{u,k})(Y_{u,k}^{(i)} Z_{u,k})^{T} \bigg]{\bigg[\rho X_{u,k} + \alpha_u\sum_{i=1}^{I_u} {A_{u,k+1}^{(i)} (Y_{u,k}^{(i)} Z_{u,k})^{T}}\bigg]^{\dagger}}, u\in\Xi(3),
\end{equation}
\begin{equation}\label{eq:4.6}
{Z}_{u,k+1} = \bigg[\rho I+\alpha_u\sum_{i=1}^{I_u}(X_{u,k+1} Y_{u,k+1}^{(i)})^{T}(X_{u,k+1} Y_{u,k+1}^{(i)}) \bigg]^{\dagger}{\bigg[\rho Z_{u,k} + \alpha_u\sum_{i=1}^{I_u} {(X_{u,k+1} Y_{u,k}^{(i)})^{T}}A_{u,k+1}^{(i)}\bigg]}, u\in\Xi(3),
\end{equation}
It's clear that (\ref{eq:4.2}) and (\ref{eq:4.4}) are both strictly convex optimization problems with strongly convex objective function, which possess global and unique minimizers.
\par (\ref{eq:4.2}) is equivalent to the following unconstrained problem
\begin{equation}\label{eq:4.7}
Y_{u,k+1}^{(i)}=\mathop {\arg \min }\limits_{Y_u^{(i)}} \dfrac{1}{2}\Vert A_{u,k+1}^{(i)}-X_{u,k+1} Y_u^{(i)} Z_{u,k}\Vert_F^2+\frac{\rho}{2}\left\|Y_u^{(i)}-Y_{u,k}^{(i)}\right\|_F^2, i\in\Xi(I_u)
\end{equation}
\noindent Treating $ Y_u^{(i)} $ as variable of the objective function in the above problem, it's then easy to check that the unique solution of the above problem is actually the solution of the following matrix equation
\begin{equation}\label{eq:4.8}
    \alpha_u X_{u,k+1}^T X_{u,k+1} Y_u^{(i)} Z_{u,k} Z_{u,k}^T+\rho Y_u^{(i)}=\Gamma^{(i)}
\end{equation}
\noindent where $ \Gamma^{(i)}=\alpha_u X_{u,k+1}^T A_{u,k+1}^{(i)} Z_{u,k}^T+\rho Y_{u,k}^{(i)} $. Notice that $ X_{u,k+1}^T X_{u,k+1} $ and $ Z_{u,k} Z_{u,k}^T $ are symmetric matrices, then there exist orthogonal matrices $ Q_1$ and $Q_2$, such that $ X_{u,k+1}^T X_{u,k+1}=Q_1 \Lambda_1 Q_1^T $ and $ Z_{u,k} Z_{u,k}^T=Q_2 \Lambda_2 Q_2^T $, where $ \Lambda_1 $ and $\Lambda_2$ are diagonal matrices whose diagonal elements are eigenvalues of $X_{u,k+1}^T X_{u,k+1}$ and $Z_{u,k} Z_{u,k}^T$, respectively. Multiplying $ Q_1^T $ from the left and multiplying $ Q_2 $ from the right on both sides of (\ref{eq:4.8}), we can get
\begin{equation}\label{eq:4.9}
    \alpha_u \Lambda_1 \hat{Y}_u^{(i)} \Lambda_2 + \rho \hat{Y}_u^{(i)} = \hat{\Gamma}^{(i)}
\end{equation}
\noindent where $ \hat{Y}_u^{(i)}=Q_1^T Y_u^{(i)} Q_2 $, $ \hat{\Gamma}^{(i)}=Q_1^T \Gamma^{(i)} Q_2 $. Since $\Lambda_1$ and $\Lambda_2$ are both diagonal matrices, $ \hat{Y}_u^{(i)} $ can be fast solved by
\begin{equation}\label{eq:4.10}
    \hat{Y}_u^{(i)}(m,n) = \dfrac{\hat{\Gamma}^{(i)}(m,n)}{\rho+\alpha_u \Lambda_1(m)\Lambda_2(n)}.
\end{equation}
Hense, $ Y_{u,k+1}^{(i)} $ in (\ref{eq:4.7}) can be solved by
\begin{equation}\label{eq:4.11}
     Y_u^{(i)} = Q_1 \hat{Y}_u^{(i)} Q_2^T.
\end{equation}
Note that
\begin{equation*}
  \begin{split}
    &\quad\sum_{u=1}^3 \alpha_u \Vert{\cal A}-\llbracket{\cal X}_u;{\cal Y}_u;{\cal Z}_u\rrbracket_u\Vert_F^2 \\
    &=\sum_{u=1}^3 \alpha_u \left\langle{\cal A}-\llbracket{\cal X}_u;{\cal Y}_u;{\cal Z}_u\rrbracket_u,{\cal A}-\llbracket {\cal X}_u;{\cal Y}_u;{\cal Z}_u\rrbracket_u \right\rangle \\
    &=\sum_{u=1}^3 \alpha_u\left\langle\mathcal{A},\mathcal{A}\right\rangle-2\sum_{u=1}^3\alpha_u\left\langle\mathcal{A},\llbracket{\cal X}_u;{\cal Y}_u;{\cal Z}_u\rrbracket_u\right\rangle+\sum_{u=1}^3 \alpha_u\left\langle\llbracket{\cal X}_u;{\cal Y}_u;{\cal Z}_u\rrbracket_u,\llbracket{\cal X}_u;{\cal Y}_u;{\cal Z}_u\rrbracket_u\right\rangle \\
    &=\left\langle\mathcal{A},\mathcal{A}\right\rangle-2\left\langle\mathcal{A},\sum_{u=1}^3\alpha_u\llbracket{\cal X}_u;{\cal Y}_u;{\cal Z}_u\rrbracket_u\right\rangle+\sum_{u=1}^3 \alpha_u\Vert\llbracket{\cal X}_u;{\cal Y}_u;{\cal Z}_u\rrbracket_u\Vert_F^2\\
    &=\left\langle\mathcal{A}-\sum_{u=1}^3\alpha_u\llbracket{\cal X}_u;{\cal Y}_u;{\cal Z}_u\rrbracket_u,\mathcal{A}-\sum_{u=1}^3\alpha_u\llbracket{\cal X}_u;{\cal Y}_u;{\cal Z}_u\rrbracket_u\right\rangle+\sum_{u=1}^3 \alpha_u\Vert\llbracket{\cal X}_u;{\cal Y}_u;{\cal Z}_u\rrbracket_u\Vert_F^2-\Vert\sum_{u=1}^3\alpha_u\llbracket{\cal X}_u;{\cal Y}_u;{\cal Z}_u\rrbracket_u\Vert_F^2 \\
    &=\Vert\mathcal{A}-\sum_{u=1}^3\alpha_u\llbracket{\cal X}_u;{\cal Y}_u;{\cal Z}_u\rrbracket_u\Vert_F^2+\sum_{u=1}^3 \alpha_u\Vert\llbracket{\cal X}_u;{\cal Y}_u;{\cal Z}_u\rrbracket_u\Vert_F^2-\Vert\sum_{u=1}^3\alpha_u\llbracket{\cal X}_u;{\cal Y}_u;{\cal Z}_u\rrbracket_u\Vert_F^2,
  \end{split}
\end{equation*}
(\ref{eq:4.4}) can be rewritten as
\begin{equation}\label{eq:4.12}
\mathcal{A}_{k+1}=\mathop {\arg \min }\limits_{\mathcal{A}} \frac{1}{2}\Vert \mathcal{A}-\sum_{u=1}^3\alpha_u\llbracket{\cal X}_{u,k};{\cal Y}_{u,k};{\cal Z}_{u,k}\rrbracket_u\Vert_F^2+\frac{\mu}{2}\Vert\mathcal{D}_w(\mathcal{A})\Vert_F^2+\frac{\rho}{2}\Vert\mathcal{A}-\mathcal{A}_{k}\Vert_F^2.
\end{equation}
It can be solved by the following linear system:
\begin{equation}\label{eq:4.13}
  \left[(1+\rho)I+\mu\mathcal{D}_w^*\mathcal{D}_w\right]\mathcal{A}=\sum_{u=1}^3\alpha_u\llbracket{\cal X}_{u,k};{\cal Y}_{u,k};{\cal Z}_{u,k}\rrbracket_u+\rho\mathcal{A}_k,
\end{equation}
where $\mathcal{D}_w^*$ denotes the adjoint operator of $\mathcal{D}_w$. Finally, by the 3D Fourier transform (fftn) and its inverse transform (ifftn), we can obtained the closed-form solution
\begin{equation*}
  \tilde{\mathcal{A}}_{k+1}=\mbox{ifftn}\left(\dfrac{\mbox{fftn}(\sum_{u=1}^3\alpha_u\llbracket{\cal X}_{u,k};{\cal Y}_{u,k};{\cal Z}_{u,k}\rrbracket_u+\rho\mathcal{A}_k)}{(1+\rho)\textbf{1}+\mu (\left|\mbox{fftn}(\mathcal{D}_h)\right|^2+\left|\mbox{fftn}(\mathcal{D}_v)\right|^2+\left|\mbox{fftn}(\mathcal{D}_t)\right|^2)}\right).
\end{equation*}
Therefore, $\mathcal{A}_{k+1}$ can be solved by
\begin{equation} \label{eq:4.20}
\mathcal{A}_{k+1}(i,j,s)=\left\{
               \begin{array}{ll}
                 \min\{1,\max\{\tilde{\mathcal{A}}_{k+1}(i,j,s),0\}\}, & (i,j,s)\notin\Omega, \\
                 \mathcal{M}(i,j,s), & (i,j,s)\in\Omega.
               \end{array}
             \right.
\end{equation}

Algorithm \ref{alg:A} presents the PAM algorithm for solving (\ref{eq:3.4}).
\begin{algorithm}[H]
\renewcommand{\thealgorithm}{1} 
\caption{The MTTD3R algorithm to solve (\ref{eq:3.4})}
\label{alg:A}
\begin{algorithmic}
\STATE {\textbf{Input:} The observed data $\mathcal{M} \in \mathbb{R}^{I_1\times I_2\times I_3}$, index set $\Omega$.}
\STATE {\textbf{Parameters:} The initialized MTT ranks $(r_1^u, r_2^u)$, $\alpha_u$, $w$, $\mu\geq 0$, $\rho =5e-5$.}
\STATE {\textbf{Initialize:} Construct ${\cal X}_{u,0}$, $\mathcal{Y}_{u,0}$, ${\cal Z}_{u,0}$ using TT-SVD\cite{ko2020fast} of $\vec{\mathcal{M}}^u$, $\epsilon = 1e-6$, iteration N=500.}
\WHILE {\textbf{$k\leq N$ and not converged}}
\STATE  (1) Fix ${\cal X}_{u,k}$, $\mathcal{Y}_{u,k}$, ${\cal Z}_{u,k} $, $u\in\Xi(3)$ to update ${\cal A}_{k+1}$ via (\ref{eq:4.20}).
\STATE  (2) Fix ${\cal A}_{k+1}$, ${\cal Y}_{u,k}$ and $\mathcal{Z}_{u,k}$ to update ${\cal X}_{u,k+1}$ via (\ref{eq:4.5}).
\STATE  (3) For every $n\in\Xi(I_u)$, fix ${\cal A}_{k}^{(n)}$, $\mathcal{X}_{u,k+1}$ and $\mathcal{Z}_{u,k}$ to update ${\cal Y}_{u,k+1}^{(n)}$ via (\ref{eq:4.7}).
\STATE  (4) Fix $\mathcal{A}_{k+1}$, $\mathcal{X}_{u,k+1}$, $\mathcal{Y}_{u,k+1}$ to update ${\cal Z}_{u,k+1}$ via (\ref{eq:4.6}).
\STATE  (5) Check the stopping criterion:$\frac{\|\mathcal{A}^{k+1}-\mathcal{A}^{k}\|_F^2}{\|\mathcal{A}^{k+1}\|_F^2}\leq \epsilon$.
\ENDWHILE
\STATE{\textbf{Output :} Recovered tensor $\mathcal{A}^{k+1}$.}
\end{algorithmic}
\end{algorithm}

\subsection{Convergence analysis}
In this section, we will prove the global convergence of Algorithm \ref{alg:A}. For convenience, we rewrite the objective function (\ref{eq:3.4}) as
\begin{equation}\label{eq:4.21}
    f(\boldsymbol{x}, \boldsymbol{y}, \boldsymbol{z}, \mathcal{A}) = h(\boldsymbol{x}, \boldsymbol{y}, \boldsymbol{z}, \mathcal{A}) + g (\mathcal{A})+\delta_S(\mathcal{A}),
\end{equation}
\noindent where $\boldsymbol{x} = ({\cal X}_1; {\cal X}_2; {\cal X}_3)$, $\boldsymbol{y} = ({\cal Y}_1; {\cal Y}_2; {\cal Y}_3)$, $\boldsymbol{z} = ({\cal Z}_1; {\cal Z}_2; {\cal Z}_3)$, $ h=f_1+f_2+f_3 $.
 \par To show the convergence of PAM algorithm, the following convergence theory is needed.
\begin{lemma}{\cite{Attouch2013}}\label{lem1}
	Let $ f:\mathbb{R}^n\rightarrow\mathbb{R}\cup\left\lbrace +\infty\right\rbrace $ be a $\mbox{PLSC}$ function. Let $ \left\lbrace \boldsymbol{x}_k\right\rbrace_{k\in\mathbb{N}}\subset\mathbb{R}^n  $ be a sequence such that
	\begin{itemize}
		\item[\rm{\textbf{H1}}] $ \left( \mbox{Sufficient decrease condition}\right) $ For each $ k\in\mathbb{N} $, there exits $ a\in\left( 0,+\infty\right) $ such that $ f\left(\boldsymbol{x}_{k+1}\right)+a\Vert\boldsymbol{x}_{k+1}-\boldsymbol{x}_k\Vert_2^2\leq f\left(\boldsymbol{x}_k\right)  $ hold；
		\item[\rm{\textbf{H2}}] $ \left( \mbox{Relative error condition}\right) $ For each $ k\in\mathbb{N} $, there exits $\boldsymbol{w}_{k+1}\in\partial f\left(\boldsymbol{x}_{k+1}\right) $ and a constant $ b\in\left( 0,+\infty\right)  $ such that $\Vert \boldsymbol{w}_{k+1}\Vert_2\leq b\Vert\boldsymbol{x}_{k+1}-\boldsymbol{x}_k\Vert_2 $ hold；
		\item[\rm{\textbf{H3}}] $ \left( \mbox{Continuity condition}\right) $There exists a subsequence $ \left\lbrace \boldsymbol{x}_{k_j}\right\rbrace_{j\in\mathbb{N}} $ and $ \bar{\boldsymbol{x}}\in\mathbb{R}^n $ such that
		\begin{equation*}
			\boldsymbol{x}_{k_j}\rightarrow\bar{\boldsymbol{x}}\mbox{ and }f\left(\boldsymbol{x}_{k_j}\right)\rightarrow f\left( \bar{\boldsymbol{x}}\right),j\rightarrow\infty.
		\end{equation*}
	\end{itemize}
	If $f$ has the K{\L} property at $ \bar{\boldsymbol{x}} $, then
	\begin{enumerate}[(i)]
		\item $\boldsymbol{x}_k\rightarrow\bar{\boldsymbol{x}} $
		\item $ \bar{\boldsymbol{x}} $ is a critical point of $f$, i.e., $ 0\in\partial f\left( \bar{\boldsymbol{x}}\right)  $;
		\item the sequence $ \left\lbrace\boldsymbol{x}_k\right\rbrace_k\in\mathbb{N} $ has a finite length, i.e.,
		\begin{equation*}
			\sum_{k=0}^{+\infty}\Vert\boldsymbol{x}_{k+1}-\boldsymbol{x}_k\Vert_2<+\infty.
		\end{equation*}
	\end{enumerate}
\end{lemma}
\par Next, we show that the objective function $ f $ in (\ref{eq:4.21}) and the sequence $ \left(\boldsymbol{x}_k, \boldsymbol{y}_k, \boldsymbol{z}_k, \mathcal{A}_k \right) $ generated by PAM algorithm satisfy the assumptions in \textbf{Lemma \ref{lem1}}. Hence, we establish the following convergence theorem.

\begin{theorem}
    Assume that the sequence $ \left(\boldsymbol{x}_k, \boldsymbol{y}_k, \boldsymbol{z}_k, \mathcal{A}_k \right) $ generated by Algorithm \ref{alg:A} is bounded. Then, the algorithm can converge to a critical point of $ f $.
\end{theorem}
\noindent \textbf{Proof.}
		First, we need to prove that $f$ is a proper lower semi-continuous function. $S$ is a non-empty closed set, which means that $\delta_S(\cdot)$ is a proper lower semi-continuous (PLSC) function. Moreover, we can know that $h$ is a polynomial function with respect to $(\boldsymbol{x}, \boldsymbol{y}, \boldsymbol{z}, \mathcal{A})$ by the definition of Frobenius norm. Then $h$ is semi-algebraic and thus a K{\L} function (which is intrinsically PLSC)\cite{Attouch2013}. Similarly, $g$ is also lower semi-continuous and proper, so the function $f$ is proper semi-continuous function.

\par Second, it's easy to see that Algorithm \ref{alg:A} is  an example of algorithm (61)-(63) displayed in \cite{Attouch2013} with $B_i=\rho I$. Thus, the sequence $ \left(\boldsymbol{x}_k, \boldsymbol{y}_k, \boldsymbol{z}_k, \mathcal{A}_k \right) $ generated by Algorithm \ref{alg:A} satisfy the conditions \textbf{H1}, \textbf{H2}, \textbf{H3} in Lemma \ref{lem1}.

\par Third, we show that $f$ satisfies the KL property at each $ \left(\boldsymbol{x}_k, \boldsymbol{y}_k, \boldsymbol{z}_k, \mathcal{A}_k \right) $, that is, $f$ is semi-algebraic on $dom(f)$. On $S$, $f$ can be expressed as $f\coloneqq h(\boldsymbol{x}, \boldsymbol{y}, \boldsymbol{z}, \mathcal{A})+g({\cal A}),\ \mathcal{A}\in S$. Since finite sums and finite products of semi-algebraic functions are semi-algebraic \cite{Attouch2013}, we only need to prove that $h(\boldsymbol{x}, \boldsymbol{y}, \boldsymbol{z}, \mathcal{A})$ and $g(\mathcal{A})$ are semi-algebraic. Obviously, $h(\boldsymbol{x}, \boldsymbol{y}, \boldsymbol{z}, \mathcal{A})$ is semi-algebraic since it's a polynomial function. The function $g(\mathcal{A})$ is a finite linear combination of the absolute value function and linear polynomials, which are both semi-algebraic. Therefore, $f$ is semi-algebraic.
\par According to Lemma \ref{lem1}, the bounded sequences generated by Algorithm \ref{alg:A} converge to a critical point of $f$ . Therefore, the proof is complete.

\section{Numerical experiments}\label{section:5}
To evaluate the performance of our MTTD3R method, extensive experiments are conducted on real-world visual data, such as color images, gray videos, MSIs and HSIs. To facilitate the numerical calculation and visualization, all testing datasets are normalized to $\left[0, 1\right]$ and they will be stretched to the original level after recovery. The compared methods include $\mbox{TCTF}$\cite{Zhou2018Tensor}, $\mbox{LRTC-TV-I}$\cite{li2017low}, $\mbox{TTD2R}$, $\mbox{TMac-dec}$\cite{Xu2013Parallel}, $\mbox{TMac-inc}$\cite{Xu2013Parallel}, and $\mbox{HaLRTC}$\cite{Liu2013Tensor}, where $\mbox{TTD2R}$ is our latest work and based on TT factorization and smoothness along only two modes. Parameters of all methods are set based on authors' codes or suggestions in their articles. Since MTTD3R is a multi-mode extension of TTD2R, we choose the same initial ranks, smoothness regularization parameters and proximal term parameters for fairness. All numerical experiments are implemented on Windows 10 64-bit and MATLAB R2017a running on a desktop equipped with an AMD Ryzen 7 4800H CPU with 2.90 GHz and 16 GB of RAM. For our MTTD3R method, the stopping criteria is set as $\dfrac{\left\|\mathcal{A}_{k+1}-\mathcal{A}_k\right\|_F}{\left\|P_{\Omega}(\mathcal{M})\right\|_F} \leq 10^{-6}$ and the maximum iteration number is set to be 500.

\subsection{Color images}
In this subsection, we test the proposed method on four popular color images \cite{li2017low} of size 256 $\times$ 256 $\times$ 3, named ``Airplane", ``Barbara", ``Sailboat" and ``House". It's worth noting that color images only exhibit low-rankness along the channel mode and smoothness along both spatial modes. The parameters of MTTD3R are setted as $\left(\alpha_1, \alpha_2, \alpha_3\right)=\left(0, 0, 1\right)$, $w=\left(w_1,w_2,w_3\right)=\left(1,1,0\right)$, $\mu=0.05$, $\rho=5\times 10^{-6}$. The initial MTT rank is shown in Table \ref{tab:1}.
\begin{table}[H]
  \centering
  \normalsize 
  \begin{spacing}{1.25} 
  \resizebox{0.5\textwidth}{!}{ 
  \begin{tabular}{ccccc}
     \hline
     Rank & Airplane & Barbara & Sailboat & House \\
     \cline{1-5}
     $(r_1^1,r_2^1)$ & $(3,37)$ & $(3,59)$ & $(3,62)$ & $(3,34)$ \\
     $(r_1^2,r_2^2)$ & $(38,3)$ & $(58,3)$ & $(64,3)$ & $(35,3)$ \\
     $(r_1^3,r_2^3)$ & $(37,38)$ & $(59,58)$ & $(62,64)$ & $(34,35)$ \\
     \hline
   \end{tabular}
   }
   \end{spacing}
  \caption{The initial MTT rank of four color images for MTT3R.}\label{tab:1}
\end{table}

The quality of the recovered color images is measured by the famous Peak Signal-to-Noise Ratio (PSNR)\cite{chen2015fractional} and the structural similarity index (SSIM)\cite{liu2014generalized}, which are defined by
\begin{equation*}
    \mbox{PSNR} = 10\cdot\log_{10}\dfrac{I_1 I_2 I_3 \Vert\mathcal{A}_{true}\Vert_{\infty}^2}{\left\|\mathcal{A}-\mathcal{A}_{true}\right\|_F^2},
\end{equation*}
and
\begin{equation*}
    \mbox{SSIM} = \dfrac{(2\mu_{\mathcal{A}}\mu_{\mathcal{A}_{true}})(2\sigma_{\mathcal{A}\mathcal{A}_{true}}+c_2)}{(\mu_{\mathcal{A}}^2
    \mu_{\mathcal{A}_{true}}^2+c_1)(\sigma_{\mathcal{A}}^2+\sigma_{\mathcal{A}_{true}}^2+c_2)},
\end{equation*}
where $\mathcal{A}_{true}$ is the true tensor, $\mathcal{A}$ is the recovered tensor and $N$ denotes the total number of pixels in the image; $\mu_\mathcal{A}$ and $\mu_{\mathcal{A}_{true}}$ are the mean values of images $\mathcal{A}$ and $\mathcal{A}_{true}$, $\sigma_{\mathcal{A}}$ and $\sigma_{\mathcal{A}_{true}}$ are the standard variances of $\mathcal{A}$ and $\mathcal{A}_{true}$, $\sigma_{\mathcal{A}\mathcal{A}_{true}}$ is the covariance of $\mathcal{A}$ and $\mathcal{A}_{true}$, and $c_1$ and $c_2>0$ are constants. Higher $\mbox{PSNR}$ and $\mbox{SSIM}$ values imply better image quality.


\begin{figure}[H]
  \centering
  \includegraphics[width=0.96\linewidth]{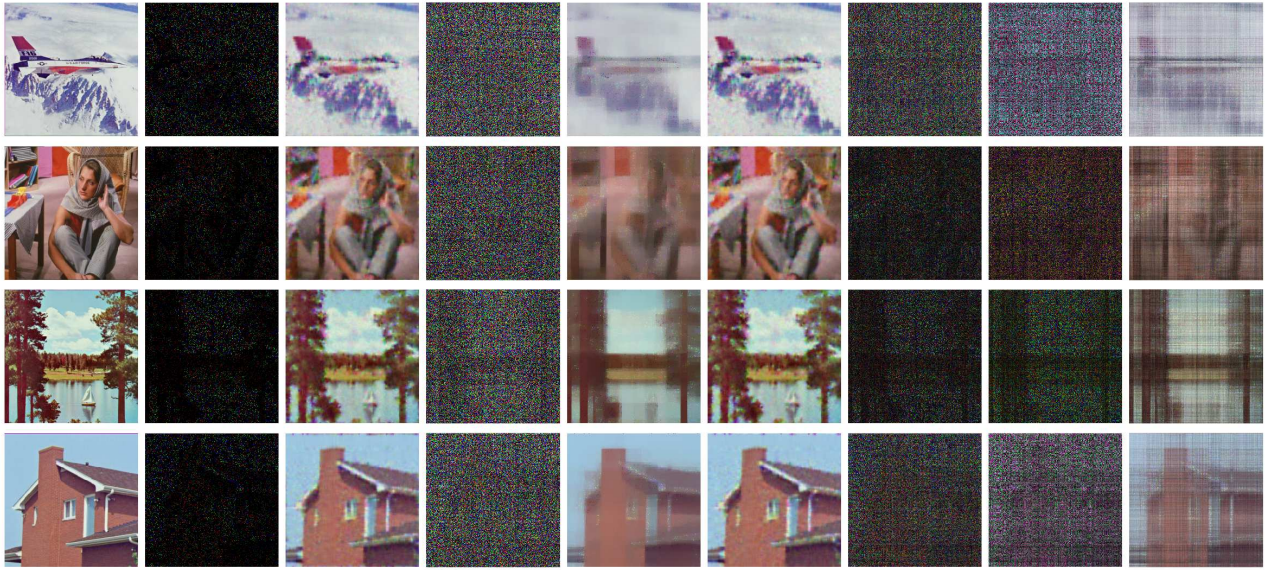}\\
  \setcaptionwidth{0.95\linewidth}
  \caption{Restored results of color images with sampling rate 10\%. From top to bottom, Airplane, Barbara, Sailboat, and House. From left to right: the original data, the observed data, the recovered results by MTTD3R, TCTF, LRTC-TV-I, TTD2R, TMac-dec, TMac-inc, and HaLRTC, respectively.}\label{fig:5}
\end{figure}

\begin{table}[htbp]
  \centering
  \tiny
  \setlength{\tabcolsep}{0.5mm}{
  \begin{tabular}{ccccccccc}
     \toprule
     Color image & p & MTTD3R & TCTF & LRTC-TV-I & TTD2R & Tmac-dec & Tmac-inc & HaLRTC \\
     \midrule
     \multirow{3}{*}{Airplane} & 0.05 & 21.85/0.68/8.97	&	4.98/0.01/2.73	&	18.98/0.61/61.50	&	\textbf{21.87}/\textbf{0.69}/11.11	&	4.09/0.01/1.39	&	5.63/0.03/1.07	&	17.09/0.38/28.64 \\
     & 0.10 & \textbf{23.81}/\textbf{0.75}/9.07	&	6.08/0.02/2.83	&	21.84/0.75/34.35	&	23.78/0.75/10.76	&	6.30/0.02/1.35	&	9.38/0.05/1.15	&	19.52/0.53/21.21 \\
     & 0.15 & 25.10/0.79/9.13	&	7.64/0.03/2.43	&	23.53/0.81/34.52	&	\textbf{25.14}/\textbf{0.79}/11.04	&	9.21/0.06/1.39	&	15.25/0.19/1.14	&	21.27/0.63/19.15 \\
     \cline{2-9}
     \multirow{3}{*}{Barbara} &	0.05	&	22.19/0.75/10.37	&	7.70/0.02/2.72	&	17.88/0.63/36.57	&	\textbf{22.36}/\textbf{0.75}/12.29	&	7.09/0.02/1.67	&	6.65/0.05/1.15	&	15.70/0.49/26.33 \\
&	0.10	&	24.09/0.81/10.48	&	8.19/0.03/2.73	&	21.69/0.76/35.49	&	\textbf{24.30}/\textbf{0.82}/12.37	&	7.98/0.04/1.70	&	9.99/0.05/1.22	&	18.69/0.63/20.45 \\
&	0.15	&	25.47/0.85/11.02	&	9.16/0.06/3.36	&	23.86/0.83/40.14	&	\textbf{25.73}/\textbf{0.85}/14.52	&	9.17/0.08/1.91	&	12.75/0.27/1.34	&	20.78/0.71/18.17 \\
\cline{2-9}
\multirow{3}{*}{Sailboat} &	0.05	&	19.62/0.73/11.19	&	6.75/0.02/2.82	&	17.25/0.63/37.39	&	\textbf{19.76}/\textbf{0.74}/12.09	&	5.77/0.03/1.78	&	4.90/0.01/1.30	&	15.23/0.37/30.81 \\
&	0.10	&	21.54/0.81/11.84	&	7.36/0.03/3.11	&	19.76/0.76/39.89	&	\textbf{21.64}/\textbf{0.81}/13.57	&	6.54/0.08/1.96	&	8.20/0.12/1.43	&	17.80/0.58/23.37 \\
&	0.15	&	22.99/0.85/11.91	&	8.65/0.06/3.06	&	21.69/0.83/40.29	&	\textbf{23.16}/\textbf{0.86}/13.76	&	7.49/0.14/2.01	&	10.00/0.23/1.48	&	19.47/0.69/19.79 \\
\cline{2-9}
\multirow{3}{*}{House}&	0.05	&	23.37/0.85/9.47	&	6.59/0.02/3.08	&	19.99/0.77/40.54	&	\textbf{23.68}/\textbf{0.86}/13.27	&	6.06/0.03/1.50	&	7.55/0.04/1.20	&	17.20/0.53/30.17 \\
&	0.10	&	25.66/0.90/10.26	&	7.46/0.02/3.17	&	23.17/0.87/39.68	&	\textbf{25.78}/\textbf{0.90}/11.19	&	8.07/0.10/1.29	&	10.97/0.22/1.06	&	20.51/0.74/24.26 \\
&	0.15	&	\textbf{27.03}/0.92/8.48	&	8.59/0.06/2.68	&	25.23/0.91/35.62	&	27.02/\textbf{0.93}/11.03	&	10.71/0.26/1.33	&	15.22/0.48/1.14	&	22.58/0.82/20.94 \\
     \bottomrule
   \end{tabular}
   }
   \setcaptionwidth{0.9\textwidth}
  \caption{The PSNR/SSIM/Running time of different methods on color images ``Airplane", ``Barbara", ``Sailboat" and ``House" with different sampling rates.}\label{tab:2}
\end{table}

The original images, the observed images and the recovered images by TCTF, TMac-inc, TMac-dec, HaLRTC, LRTC-TV-I, TTD2R and MTTD3R are displayed in Figure \ref{fig:5}. The PSNR and SSIM values of the recovered images are summarized in Tables \ref{tab:2}. Figure \ref{fig:5} and Table \ref{tab:2} show that (i) MTTD3R and TTD2R have overall better performance in term of visual visual quality among seven methods, and MTTD3R is more efficient than TTD2R in term of running time; (ii) there is little difference between the recovered results of MTTD3R and TTD2R, which means that they are essentially equivalent, only the iteration order is different and the smoothness terms are slightly different; (iii) when sampling rate is low, TCTF and TMac (-inc and -dec) cannot accurately restore the incomplete color images.

\subsection{Gray video}
In this subsection, we compare the performance of the proposed method and other methods on videos. We test 3 videos, including \emph{hall} of size $144\times 176\times 300$, \emph{suize} of size $144\times 176\times 150$ and \emph{salesman} of size $144\times 176\times 449$\footnote{http://trace.eas.asu.edu/yuv/}. We use 30 frames of \emph{hall} and \emph{suize}, thus both test tensors are of size $144\times 176\times 30$. For \emph{salesman} video, we choose 50 frames, that is, the test tensor is of size $144\times 176\times 50$. The parameters of our model are setted as $\left(\alpha_1, \alpha_2, \alpha_3\right)=\left(\frac{1}{3}, \frac{1}{3}, \frac{1}{3}\right)$, $w=\left(w_1,w_2,w_3\right)=\left(1,1,1\right)$, $\mu=0.005$, $\rho=5\times 10^{-6}$. The initial MTT rank is shown in Table \ref{tab:5}.

\begin{table}[H]
  \centering
  \small
  \begin{spacing}{1.25} 
  \resizebox{0.4\textwidth}{!}{ 
  \begin{tabular}{cccc}
    \hline
    Video & $(r_1^1,r_2^1)$ & $(r_1^2,r_2^2)$ & $(r_1^3,r_2^3)$ \\
    \cline{1-4}
    \emph{hall} & (7,41) & (40,7) & (41,40) \\
    \emph{suize} & (9,31) & (37,9) & (31,37) \\
    \emph{salesman} & (16,76) & (72,16) & (76,72) \\
    \hline
  \end{tabular}
  }
  \end{spacing}
  \caption{The initial MTT rank of test videos for MTTD3R.}\label{tab:5}
\end{table}

\par Two quantitative indices, i.e., the mean peak signal-to-noise ratio ($\mbox{MPSNR}$) and mean structural similarity ($\mbox{MSSIM}$)\cite{wang2004image}, are used in our video experiments
\begin{equation*}
  \mbox{MPSNR}=\frac{1}{B}\sum_{i=1}^B\mbox{PSNR}_i,
\end{equation*}
\begin{equation*}
  \mbox{MSSIM}=\frac{1}{B}\sum_{i=1}^B\mbox{SSIM}_i,
\end{equation*}
where $\mbox{PSNR}_i$ and $\mbox{SSIM}_i$ are the $\mbox{PSNR}$ and $\mbox{SSIM}$ values for the $i$th frame, respectively.
\par
\begin{figure}[H]
  \centering
  \includegraphics[width=0.95\linewidth]{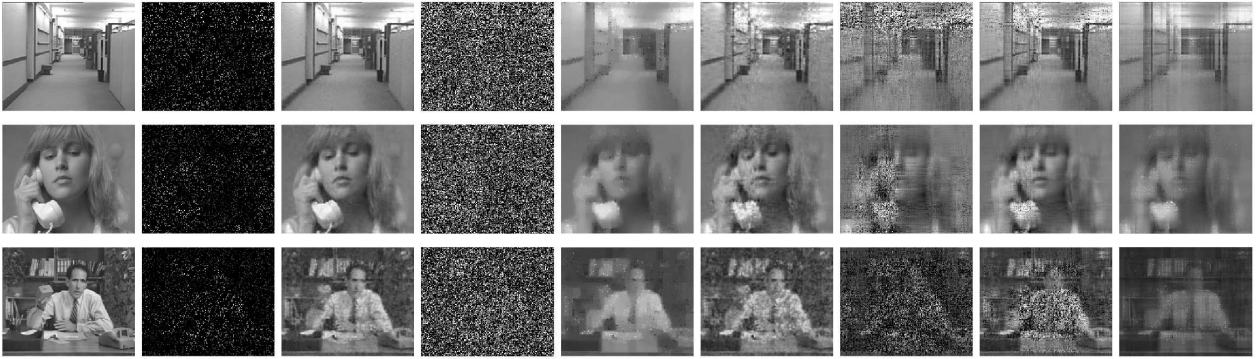}\\
  \setcaptionwidth{0.95\linewidth}
  \caption{Restored results on video recovery with 10\% observed entries. From up to down: \emph{hall}, \emph{suize} and \emph{salesman}. From left to right: 5th frame of the original data, the observed data and the recovered results by MTTD3R, TCTF, LRTC-TV-I, TTD2R, TMac-dec, TMac-inc, and HaLRTC, respectively.}\label{fig:6}
\end{figure}

 Because of the page limitation, we only present the 5th frame of the \emph{hall}, \emph{suize} and \emph{salesman} video before and after recovering in Figure \ref{fig:6}. To further evaluate the overall performance of the proposed method, we give the quantitative comparison for all experimental cases in Table \ref{tab:3}. Figure \ref{fig:6} and Table \ref{tab:3} indicate that (i) MTTD3R have the best performance in term of visual quality among the seven methods; (ii) methods with local smoothness constraint such as $\mbox{MTTD3R}$, $\mbox{TTD2R}$ and $\mbox{LRTC-TV-I}$ perform better than other methods when the sampling rate is small.

\begin{table}[htbp]
  \centering
  \tiny
  \setlength{\tabcolsep}{1mm}{

  \begin{tabular}{ccccccccc}
     \toprule
     Video & p & MTTD3R & TCTF & LRTC-TV-I & TTD2R & Tmac-dec & Tmac-inc & HaLRTC \\
     \midrule
     \multirow{3}{*}{\emph{hall}} & 0.05 & \textbf{26.58}/\textbf{0.83}/71.23	&	6.44/0.01/7.94	&	19.05/0.58/93.36	&	20.46/0.54/38.01	&	9.14/0.07/4.01	&	13.78/0.28/3.39	&	18.72/0.27/45.96 \\
     & 0.10 & \textbf{30.76}/\textbf{0.90}/71.69	&	6.96/0.02/13.55	&	21.41/0.71/92.05	&	23.78/0.73/38.49	&	13.89/0.37/4.13	&	21.90/0.68/3.45	&	22.02/0.45/26.91 \\
     & 0.15 & \textbf{32.25}/\textbf{0.92}/72.19	&	7.51/0.03/13.76	&	23.19/0.79/91.87	&	26.61/0.83/39.04	&	18.46/0.63/4.23	&	28.58/0.87/3.54	&	24.38/0.57/19.47 \\
     \cline{2-9}
     \multirow{3}{*}{\emph{suize}} &	0.05	&	\textbf{29.10}/\textbf{0.80}/69.03	&	7.67/0.01/7.90	&	21.72/0.64/93.86	&	24.13/0.64/35.83	&	11.73/0.07/3.88	&	16.73/0.26/3.20	&	20.70/0.22/43.02 \\
&	0.10	&	\textbf{31.21}/\textbf{0.85}/70.14	&	8.17/0.01/13.53	&	25.98/0.76/92.50	&	27.17/0.74/36.45	&	17.56/0.37/4.00	&	24.59/0.70/3.35	&	24.41/0.40/21.18 \\
&	0.15	&	\textbf{32.17}/\textbf{0.88}/70.26	&	8.65/0.01/13.73	&	28.16/0.82/92.02	&	29.14/0.81/36.15	&	22.02/0.65/4.04	&	29.81/0.84/3.61	&	26.56/0.51/13.88 \\
\cline{2-9}
\multirow{3}{*}{\emph{salesman}} &	0.05	&	\textbf{21.42}/\textbf{0.60}/73.91	&	7.91/0.01/16.25	&	17.95/0.38/144.67	&	18.85/0.44/86.69	&	8.60/0.04/8.56	&	11.83/0.17/6.58	&	17.52/0.26/62.64 \\
&	0.10	&	\textbf{24.48}/\textbf{0.73}/67.12	&	8.46/0.02/22.88	&	20.79/0.53/141.69	&	21.46/0.58/86.85	&	10.45/0.12/8.74	&	16.40/0.49/7.03	&	20.38/0.45/35.32 \\
&	0.15	&	\textbf{26.13}/\textbf{0.80}/67.91	&	8.95/0.03/23.20	&	22.57/0.64/140.69	&	23.25/0.67/87.45	&	12.42/0.28/9.03	&	20.88/0.71/7.46	&	22.39/0.58/25.19 \\
     \bottomrule
   \end{tabular}
   }

   \setcaptionwidth{0.9\linewidth}
  \caption{The MPSNR/MSSIM/Running time of different methods on gray videos \emph{hall}, \emph{suize} and \emph{salesman} with different sampling rates.}\label{tab:3}
\end{table}

\subsection{Multispectral image and hyperspectral remote sensing image recovery}\label{sec:rsir}
In this subsection, we conduct experiments on two HSI data (\emph{Washington DC Mall}\footnote{https://engineering.purdue.edu/~biehl/MultiSpec/hyperspectral.html} of size $256 \times 256 \times 10$ and \emph{Pavia City Center}\footnote{http://www.ehu.eus/ccwintco/index.php?title=Hyperspectral\_Remote\_Sensing\_Scenes} of size $200\times 200\times 80$) and the MSI data\footnote{https://www1.cs.columbia.edu/CAVE/databases/multispectral/} of size $256\times 256\times 31$. We still employ MPSNR and MSSIM to measure the quality of the recovered results. The parameters of our model are setted as $\left(\alpha_1, \alpha_2, \alpha_3\right)=\left(\frac{1}{3}, \frac{1}{3}, \frac{1}{3}\right)$, $w=\left(w_1,w_2,w_3\right)=\left(1,1,1\right)$, $\mu=0.01$, $\rho=5\times 10^{-6}$. The initial MTT rank is shown in Table \ref{tab:6}.

\begin{table}[H]
  \centering
  \small
  \begin{spacing}{1.25} 
  \resizebox{0.5\textwidth}{!}{ 
  \begin{tabular}{ccccc}
    \toprule
    \multicolumn{2}{c}{Data} & $(r_1^1,r_2^1)$ & $(r_1^2,r_2^2)$ & $(r_1^3,r_2^3)$ \\
    \midrule
    \multirow{2}{*}{HSI} & \emph{Washington} & (2,99) & (99,2) & (99,99) \\
    \cline{2-5}
    & \emph{Pavia} & (5,100) & (97,5) & (100,97) \\
    \midrule
    \multicolumn{2}{c}{MSI} & $(7,145)$ & $(142,7)$ & $(145,142)$ \\
    \bottomrule
  \end{tabular}
  }
  \end{spacing}
  \caption{The MTT rank of test data for MTT3R.}\label{tab:6}
\end{table}

\par
\begin{figure}[H]
  \centering
  \includegraphics[width=0.95\linewidth]{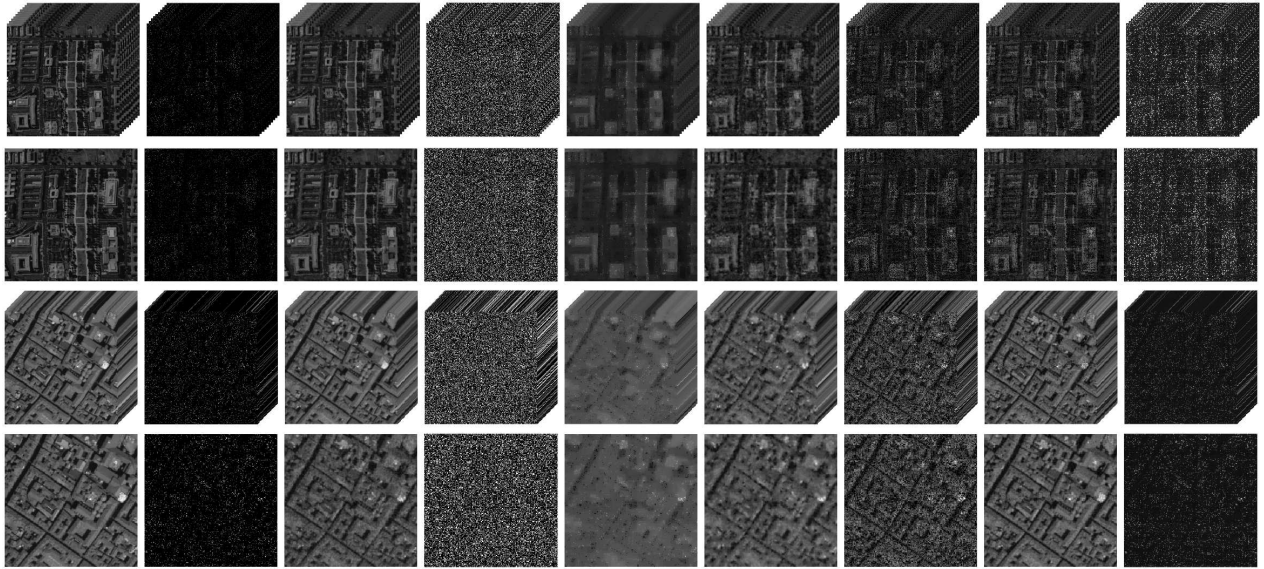}\\
  \setcaptionwidth{0.95\linewidth}
  \caption{Restored results on HSI with 10\% observed entries. From up to down: 3-D visualization and last band of HSI \emph{Washington DC Mall}, 3-D visualization and last band of HSI \emph{Pavia City Center}. From left to right: the original data, the observed data and the recovered results by MTTD3R, TCTF, LRTC-TV-I, TTD2R, TMac-dec, TMac-inc, and HaLRTC, respectively.}\label{fig:7}
\end{figure}

\begin{figure}[H]
  \centering
  \includegraphics[width=0.95\linewidth]{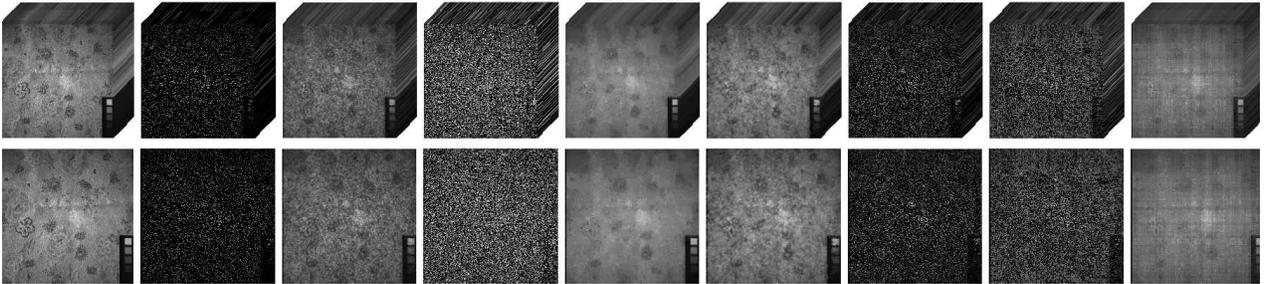}\\
  \setcaptionwidth{0.95\linewidth}
  \caption{Restored results on MSI with 10\% observed entries, 3-D visualization and the last band of MSI data. From left to right: the original data, the observed data and the recovered results by MTTD3R, TCTF, LRTC-TV-I, TTD2R, TMac-dec, TMac-inc, and HaLRTC, respectively.}\label{fig:7_1}
\end{figure}

\begin{figure}[H]
  \centering
  \includegraphics[width=0.9\linewidth]{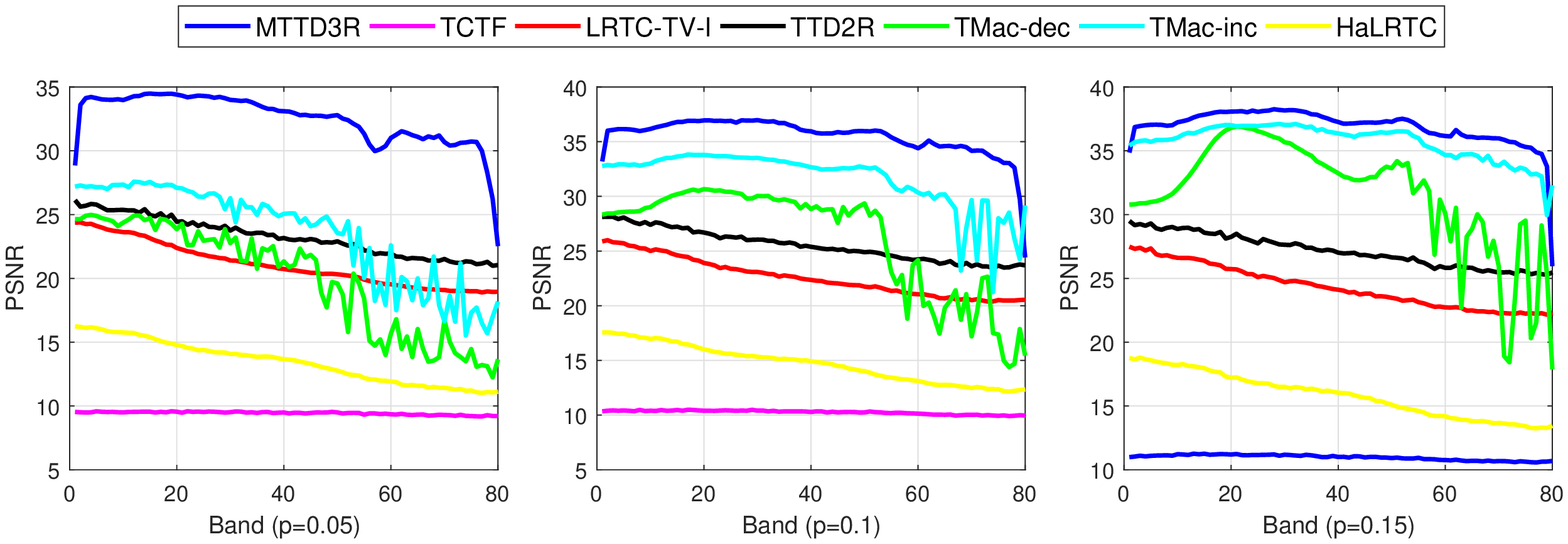}\\
  \caption{PSNR values of each band of the recovered HSI \emph{Pavia City Center}.}\label{fig:9}
\end{figure}

\begin{figure}[H]
  \centering
  \includegraphics[width=0.9\linewidth]{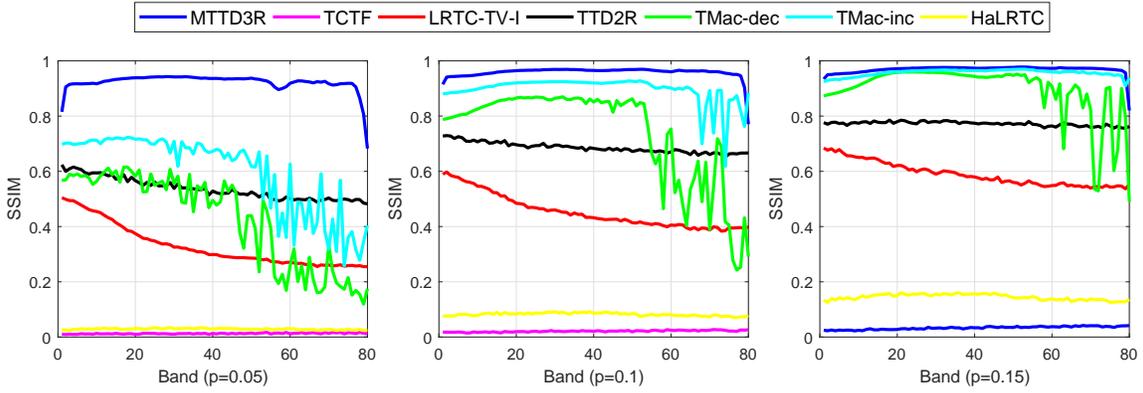}\\
  \caption{SSIM values of each band of the recovered HSI \emph{Pavia City Center}.}\label{fig:10}
\end{figure}

 Table \ref{tab:7_1} give the quantitative comparison of all recovered results on HSI and MSI, while Figure \ref{fig:7} and \ref{fig:7_1} display the visualization results. In addition, Figure \ref{fig:9} and Figure \ref{fig:10} give the PSNR and SSIM values comparison of each band of the HSI \emph{Pavia City Center} recovered by all methods, respectively. The results on the other two datasets are similar, we will not go into details because of page limitation. Table \ref{tab:7_1} and Figure \ref{fig:7}-\ref{fig:10} show that (i) MTTD3R has the best performance in term of visual quality among all methods, especially for large-scale data with low sampling rate; (ii) methods with local smoothness constraint such as $\mbox{MTTD3R}$, $\mbox{TTD2R}$ and $\mbox{LRTC-TV-I}$ perform better than other methods when the sampling rate is small. Besides, it can be seen from Table \ref{tab:7_1}, Figure \ref{fig:9} and Figure \ref{fig:10} that $\mbox{MPSNR}$ and $\mbox{MSSIM}$ can well reflect the recovered quality of each band.

\begin{table}[htbp]
  \centering
  \tiny
  \setlength{\tabcolsep}{0.5mm}{
  \begin{tabular}{cccccccccc}
     \toprule
     \multicolumn{2}{c}{Data} & p & MTTD3R & TCTF & LRTC-TV-I & TTD2R & Tmac-dec & Tmac-inc & HaLRTC \\
     \midrule
     \multirow{6}{*}{\emph{HSI}} & \multirow{3}{*}{\emph{Washington}} & 0.05 & \textbf{27.03}/\textbf{0.79}/124.07	&	9.62/0.02/9.37	&	19.71/0.34/116.66	&	22.09/0.49/49.18	&	14.08/0.11/5.71	&	15.30/0.17/4.08	&	13.60/0.02/242.50 \\
     & & 0.10 & \textbf{30.44}/\textbf{0.90}/123.81	&	10.15/0.03/9.46	&	21.88/0.48/113.26	&	24.20/0.64/49.09	&	16.58/0.28/5.77	&	20.33/0.50/4.44	&	14.33/0.06/243.00 \\
     & & 0.15 & \textbf{33.01}/\textbf{0.94}/124.38	&	12.27/0.06/8.85	&	23.69/0.61/112.61	&	25.94/0.74/49.12	&	18.81/0.44/5.83	&	25.13/0.75/4.56	&	15.03/0.11/243.36 \\
     \cline{2-10}
     & \multirow{3}{*}{\emph{Pavia}} &	0.05	&	\textbf{32.49}/\textbf{0.92}/338.22	&	9.44/0.01/40.97	&	21.07/0.33/562.14	&	23.28/0.54/246.29	&	19.99/0.43/23.25	&	23.56/0.60/18.47	&	13.47/0.03/232.97 \\
& &	0.10	&	\textbf{35.48}/\textbf{0.96}/336.03	&	10.26/0.02/49.89	&	22.60/0.45/553.90	&	25.51/0.69/247.52	&	26.13/0.74/24.01	&	31.61/0.90/18.97	&	14.69/0.08/234.27 \\
& &	0.15	&	\textbf{36.87}/\textbf{0.97}/337.79	&	10.97/0.03/50.14	&	24.32/0.59/550.38	&	27.16/0.77/248.67	&	31.62/0.90/24.58	&	35.71/0.96/20.97	&	15.85/0.14/235.82 \\
\midrule
\multicolumn{2}{c}{\multirow{3}{*}{MSI}} &	0.05	&	\textbf{21.92}/\textbf{0.64}/372.56	&	6.88/0.01/31.84	&	18.09/0.37/461.50	&	18.51/0.37/229.43	&	13.01/0.16/21.05	&	15.43/0.29/14.91	&	17.70/0.12/106.52 \\
& &	0.10	&	\textbf{24.41}/\textbf{0.77}/373.25	&	7.46/0.02/32.28	&	19.74/0.47/456.42	&	20.35/0.50/237.26	&	15.39/0.33/21.55	&	18.84/0.48/15.92	&	19.22/0.24/72.25 \\
& &	0.15	&	\textbf{26.60}/\textbf{0.84}/361.59	&	8.40/0.03/32.25	&	20.84/0.55/474.49	&	21.39/0.58/227.20	&	18.13/0.48/21.54	&	22.03/0.64/16.60	&	20.38/0.35/55.11 \\
     \bottomrule
   \end{tabular}
   }
   \setcaptionwidth{0.9\linewidth}
  \caption{The MPSNR/MSSIM/Running time of different methods on HSI and MSI with different sampling rates.}\label{tab:7_1}
\end{table}

\subsection{Remote sensing image cloud removal}
\begin{figure}[H]
  \centering
  \includegraphics[width=0.8\linewidth]{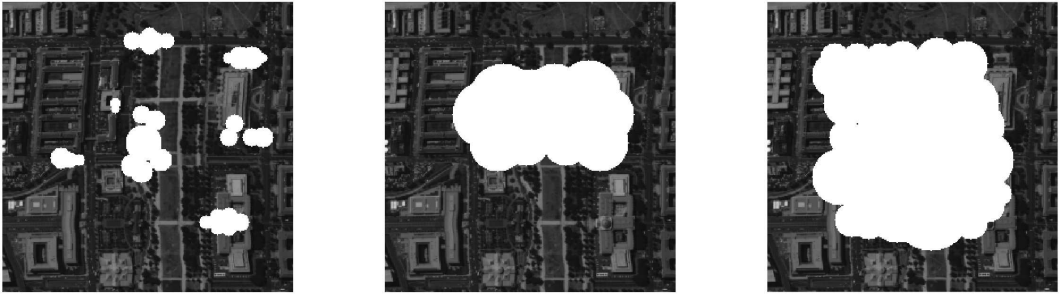}\\
  \setcaptionwidth{0.8\linewidth}
  \caption{Three cases of cloud cover. (From left to right) Case \uppercase\expandafter{\romannumeral1}: multiple small clouds. Case \uppercase\expandafter{\romannumeral2}: single middle cloud. Case \uppercase\expandafter{\romannumeral3}: single large cloud.}\label{fig:cloud}
\end{figure}

Remote sensing images are easily affected by climate factors, for example, cloud cover is one of the influencing factors. Cloud removal from remote sensing images can improve the effectiveness and availability of remote sensing data, which has very important practical significance. We conduct our experiments on a subimage of the Washington DC Mall data set of size $256 \times 256 \times 10$ and simulate three cases of cloud cover, as shown in Figure \ref{fig:cloud}. We adopt the parameter settings $\left(\alpha_1, \alpha_2, \alpha_3\right)=\left(\frac{1}{3}, \frac{1}{3}, \frac{1}{3}\right)$, $w=\left(w_1,w_2,w_3\right)=\left(1,1,1\right)$, $\mu=0.05$, $\rho=5\times 10^{-6}$ and the MTT rank is the same as in subsection \ref{sec:rsir}. Table \ref{tab:cloud_simulation} displays the experimental results of all methods. We can see that MTTD3R and LRTC-TV-I have comparable performance, but the running time of MTTD3R is half shorter than that of LRTC-TV-I. HaLRTC provides mediocre performance but takes the longest time.

\begin{table}[H]
  \centering
  \small
  \begin{spacing}{1.15} 
  \begin{tabular}{cccccccccc}
    \hline
    \multirow{2}{*}{Methods} & Case \uppercase\expandafter{\romannumeral1} & Case \uppercase\expandafter{\romannumeral2} & Case \uppercase\expandafter{\romannumeral3} \\
    \cline{2-4}
     & MPSNR/MSSIM/time & MPSNR/MSSIM/time & MPSNR/MSSIM/time \\
     \hline
     MTTD3R	& \textbf{30.37}/0.94/44.97 & \textbf{25.44}/0.85/47.96 & \textbf{22.77}/\textbf{0.69}/59.92 \\
     TCTF &	21.82/0.92/8.92	& 15.23/0.80/14.73 & 12.55/0.59/15.34 \\
     LRTC-TV-I & 30.33/\textbf{0.95}/105.96	& 24.02/\textbf{0.86}/123.82 & 20.42/\textbf{0.69}/145.92 \\
     TTD2R & 28.51/0.93/61.71	& 24.78/0.84/58.74 & 22.45/0.68/64.33 \\
     TMac-dec &	23.22/0.91/8.75	& 19.48/0.80/7.37 & 16.49/0.58/6.97 \\
     TMac-inc &	23.27/0.91/7.96	& 19.37/0.80/7.35 & 16.41/0.58/6.21 \\
     HaLRTC	& 24.97/0.92/302.74	& 20.12/0.80/281.82 & 16.93/0.58/309.25 \\
    \hline
  \end{tabular}
  \end{spacing}
  \setcaptionwidth{0.7\linewidth}
  \caption{The PSNR/SSIM/Running time of recovered remote sensing images by different methods on the \emph{Washington DC Mall}.}\label{tab:cloud_simulation}
\end{table}

\section{Conclusion}\label{section:6}
In this article, we generalize the tensor train factorization to the mode-k tensor train factorization and established a relationship between mode-k TT rank and Tucker rank. We propose a novel multi-mode TT factorization based completion model, and model visual data as the corresponding low-MTT-rank component. Then, we integrated spatial-spectral characteristics into the proposed model and obtain an improved model. We develop an efficient PAM-based algorithm with theoretical and empirical convergence. Comparing MTTD3R with the state-of-the-art completion and approximation methods such as $\mbox{TCTF}$, $\mbox{LRTC-TV-I}$, $\mbox{TCTF}$, $\mbox{TTD2R}$, $\mbox{TMac-dec}$, $\mbox{TMac-inc}$, and $\mbox{HaLRTC}$, extensive experimental results demonstrate that the proposed $\mbox{MTTD3R}$ method has superiorities of better recovering the missing entries and finely preserving the inherent structure.




\bibliographystyle{unsrt}    
\bibliography{citation} 

\end{CJK*}
\end{document}